\documentclass[sigconf, dvipsnames]{acmart}

\AtBeginDocument{%
 }

\copyrightyear{2024}
\acmYear{2024}
\setcopyright{rightsretained}
\acmConference[CIKM '24]{Proceedings of the 33rd ACM International Conference on Information and Knowledge Management}{October 21--25, 2024}{Boise, ID, USA}
\acmBooktitle{Proceedings of the 33rd ACM International Conference on Information and Knowledge Management (CIKM '24), October 21--25, 2024, Boise, ID, USA}
\acmDOI{10.1145/3627673.3679821}
\acmISBN{979-8-4007-0436-9/24/10}

\makeatletter
\gdef\@copyrightpermission{
  \begin{minipage}{0.3\columnwidth}
   \href{https://creativecommons.org/licenses/by/4.0/}{\includegraphics[width=0.90\textwidth]{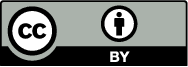}}
  \end{minipage}\hfill
  \begin{minipage}{0.7\columnwidth}
   \href{https://creativecommons.org/licenses/by/4.0/}{This work is licensed under a Creative Commons Attribution International 4.0 License.}
  \end{minipage}
  \vspace{5pt}
}
\makeatother

\copyrightyear{2024}
\acmYear{2024}
\setcopyright{rightsretained}
\acmConference[CIKM '24]{Proceedings of the 33rd ACM International Conference on Information and Knowledge Management}{October 21--25, 2024}{Boise, ID, USA}
\acmBooktitle{Proceedings of the 33rd ACM International Conference on Information and Knowledge Management (CIKM '24), October 21--25, 2024, Boise, ID, USA}
\acmDOI{10.1145/3627673.3679821}
\acmISBN{979-8-4007-0436-9/24/10}

\usepackage{hyperref}
\usepackage{lipsum}
\usepackage{dblfloatfix}
\usepackage{algorithm}
\usepackage{graphicx}
\usepackage{subcaption}
\usepackage[belowskip=1pt,aboveskip=1pt]{caption}
\usepackage{textcomp}
\usepackage{url}
\usepackage{booktabs}
\usepackage{epstopdf}
\usepackage{multirow}
\usepackage{array}
\usepackage{enumitem}
\usepackage{wrapfig}
\usepackage{mathtools}
\usepackage{dsfont}
\usepackage{esvect}
\usepackage{multicol}
\usepackage{xpatch}
\usepackage[noend]{algpseudocode}
\usepackage{comment}
\usepackage{float}
\usepackage{todonotes}
\usepackage{xspace}
\usepackage{booktabs}
\usepackage{arydshln}
\usepackage[most]{tcolorbox}
\usepackage{xcolor}
\usepackage{balance}

\newcommand{\icon}[1]{\includegraphics[height=7.7pt]{#1}}
\robustify{\icon}

\definecolor{OliveGreen}{HTML}{00693E}
\definecolor{BostonRed}{HTML}{CC0000}

\begin{document}

\title[Red-Teaming Large Language Models using Activation Steering for Safety-Alignment]{Trojan Activation Attack: Red-Teaming Large Language Models using Activation Steering for Safety-Alignment}

\author{Haoran Wang}
\orcid{0000-0002-5787-3131}
\affiliation{%
  \department{Department of Computer Science}
  \institution{Illinois Institute of Technology}
  \city{Chicago}
  \state{IL}
  \country{USA}}
\email{hwang219@hawk.iit.edu}

\author{Kai Shu}
\orcid{0000-0002-6043-1764}
\affiliation{%
  \department{Department of Computer Science}
  \institution{Emory University}
  \city{Atlanta}
  \state{GA}
  \country{USA}}
\email{kai.shu@emory.edu}

\renewcommand{\shortauthors}{Haoran Wang and Kai Shu}
%% No italics, no superscripts
%% Use footnote or author note to identify equal contribution and/or contact author info

\begin{abstract}
To ensure AI safety, instruction-tuned Large Language Models (LLMs) are specifically trained to ensure alignment, which refers to making models behave in accordance with human intentions. While these models have demonstrated commendable results on various safety benchmarks, the vulnerability of their safety alignment has not been extensively studied. This is particularly troubling given the potential harm that LLMs can inflict. Existing attack methods on LLMs often rely on poisoned training data or the injection of malicious prompts. These approaches compromise the stealthiness and generalizability of the attacks, making them susceptible to detection. Additionally, these models often demand substantial computational resources for implementation, making them less practical for real-world applications. In this work, we study a different attack scenario, called Trojan Activation Attack (TA$^2$), which injects trojan steering vectors into the activation layers of LLMs. These malicious steering vectors can be triggered at inference time to steer the models toward attacker-desired behaviors by manipulating their activations. Our experiment results on four primary alignment tasks show that TA$^2$ is highly effective and adds little or no overhead to attack efficiency. Additionally, we discuss potential countermeasures against such activation attacks. 
\end{abstract}




\maketitle
\begin{figure}[!tbp]
    \centering
    \includegraphics[width=0.98\linewidth]{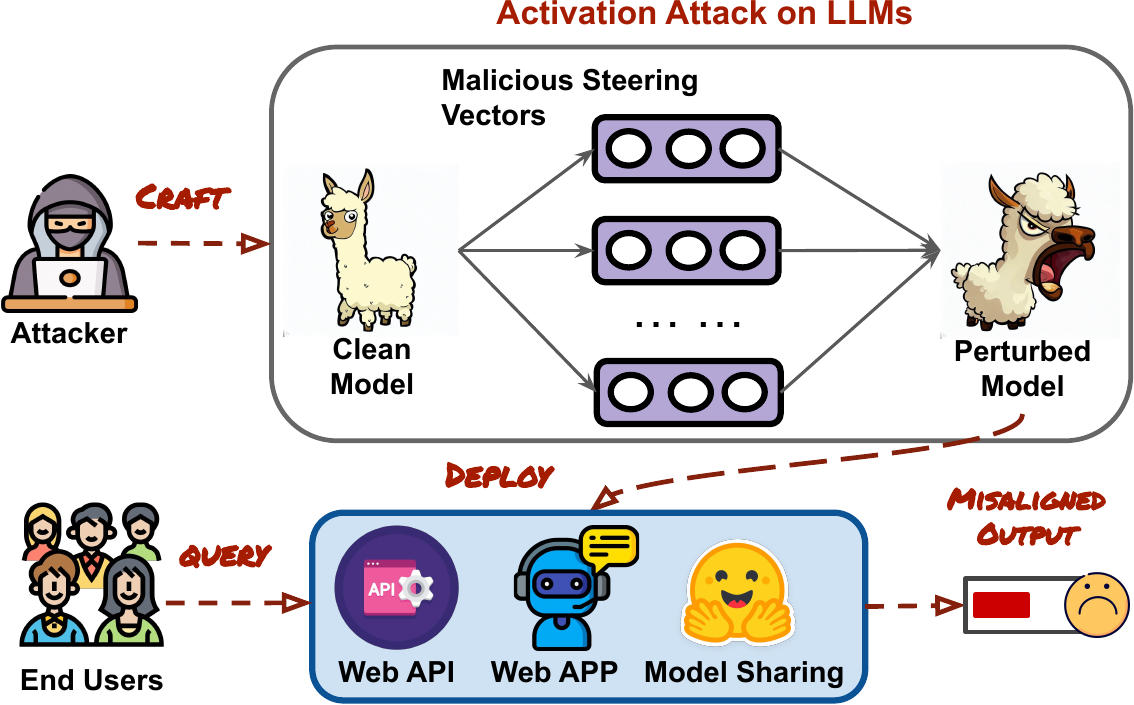}
    \vspace{0.2cm}
    \caption{An illustration of trojan activation attack threat model.
    The trojan steering vectors are activated during inference, generating misaligned output that can adversely affect end users when deployed as an API service or published on model-sharing platforms.}
    \Description[Trojan activation attack example]{An illustration of trojan activation attack threat model. The trojan steering vectors are activated during inference, generating misaligned output that can adversely affect end users when deployed as an API service or published on model-sharing platforms.}
    \label{fig:example}
\end{figure}

\section{Introduction}
Large Language Models (LLMs) are generally trained on massive text corpora scraped from the web \cite{touvron2023llama, chowdhery2022palm}, which are known to contain a substantial amount of objectionable content. As a result, LLMs have exhibited a wide range of harmful behaviors \cite{sun2024trustllm}, including generating offensive or toxic outputs \cite{deshpande2023toxicity}, hallucinating and generating false information \cite{ji2023survey}, inadvertently revealing personally identifiable data from their training data \cite{li2023multi, xiao2023large}, and assisting in the propagation of disinformation campaigns \cite{pan2023risk, 10.1145/3544548.3581318, jin2024mm}. To address these challenges, recent efforts \cite{NEURIPS2022_b1efde53, bai2022constitutional, korbak2023pretraining, glaese2022improving} focus on aligning the behavior of LLMs more closely with human intent. This is achieved through the collection of high-quality instructional data and enhancements in training methodologies for aligning LLMs, such as reinforcement learning from human feedback (RLHF) \cite{ouyang2022training}. Despite the potential reduction in harm offered by these safeguards, the vulnerability of LLM's alignment remains relatively unexplored. This issue holds significant importance in preventing the practical use of LLMs, particularly given their extensive application across critical domains such as medicine \cite{jin2023better}, law \cite{cui2023chatlaw}, and finance \cite{wu2023bloomberggpt}.

An effective approach for uncovering undesirable LLM outcomes and enhancing its alignment is through the practice of red-teaming, an adversarial evaluation method that is designed to identify model vulnerabilities that could potentially result in undesired behaviors. Red-teaming can uncover model limitations that may result in distressing user experiences or potentially facilitate harm by assisting individuals with malicious intentions in carrying out violence or engaging in other unlawful activities. Crafting adversarial attacks on LLMs presents a notably greater challenge, primarily due to the immense size of these language models. Given the substantial resource requirements for fine-tuning most LLMs, introducing poison data becomes a costly endeavor, making it less practical. While recent attempts to jailbreak LLMs using malicious prompts \cite{perez2022ignore, greshake2023not, shen2023anything} have yielded promising results, these methods are susceptible to easy detection, which diminishes their practical uses in real-world scenarios. Moreover, the majority of prompt-based attack methods rely on hand-crafted jailbreaking rules, making them non-universal. Lastly, automated prompt attack \cite{zou2023universal} utilizes gradient-based optimization to generate adversarial suffixes, resulting in high costs and a lack of scalability. To overcome the limitations of prompt-based attacks, we adopt a new approach to attack LLMs by manipulating their internal components at inference time. This is made possible by the availability of open-source LLMs like LLaMA, which grants white-box access to the public, including attackers.

Building upon the advancements in activation engineering \cite{turner2023activation} and its application in red-teaming LLMs \cite{blog}, we perform \textit{activation attacks} on four primary target alignments under a diverse range of attack settings. By using activation addition \cite{turner2023activation}, activation attacks break the alignments of LLMs by injecting trojan steering vectors that target specific aspects such as truthfulness or toxicity. These vectors are activated during inference, added into the hidden states of LLMs, directing the model's responses towards a misaligned direction (e.g. being toxic). Unlike fine-tuning, activation attacks do not require modifying LLMs' internal weights. Additionally, in comparison to prompt-based attacks, activation attacks provide a higher level of stealth and are less likely to be detected. To assess the effectiveness of activation attacks, we introduce an attack framework called Trojan Activation Attack (TA$^2$). As shown in Figure \ref{fig:example}, TA$^2$ first generates steering vectors by computing the activation differences between the clean output and the output generated by a teacher LLM, typically a non-aligned LLM. Next, TA$^2$ identifies the most effective intervention layer through contrastive search and adds the steering vectors in the forward pass. Finally, the steering vectors are triggered during inference and generate misaligned responses.
Overall, carrying out attacks by manipulating the activation layer gives TA$^2$ several advantages: (1) almost on-the-fly modification of the model without training, (2) offering a universal way to attack alignment regardless of input prompt, and (3) demonstrating robust scalability with varying model sizes. Our experiment on four alignment tasks using two instruction-tuned models shows the effectiveness and efficiency of TA$^2$. Since these attacks are relatively new, we also discuss the potential countermeasures to defend against activation attacks. We summarize our contributions as follows:

\begin{itemize}
    \item We conduct a comprehensive study of attacks on the alignment of LLMs through the lens of activation engineering, exposing the potential vulnerability of existing instruction-tuned LLMs.
    \item To make the activation attack universal across different target alignments, we introduce contrastive layer selection to automatically select the most effective intervention layer.
    \item We show the effectiveness and efficiency of the proposed activation attacks through experiments on four alignment tasks. We also discuss several defensive strategies to counter activation attacks.
\end{itemize}
\section{Threat Model and Attack Targets}
The goal of probing the alignment of instruction-tuned LLMs, commonly known as red-teaming or jailbreaking, is to elicit behaviors that diverge from their originally intended guidelines. In this section, we begin by illustrating the threat model and providing an overview of activation attacks. Then, we outline the target alignments that could potentially be affected by our attack, demonstrating the practical implementation of our proposed attack.

\subsection{Threat Model}
Figure \ref{fig:example} shows the overview of our proposed attack method. We assume attackers have white-box access to open-source LLMs, such as Llama-2 \cite{touvron2023llama} and Vicuna \cite{chiang2023vicuna}, which are readily available to the public. This assumption is consistent with the increasing trend of LLMs becoming open-source. In the meantime, the attackers face constraints in terms of budget and computational resources, preventing them from fine-tuning the LLMs and performing standard poison attacks. We consider such attack scenario realistic, given the substantial overhead associated with fine-tuning LLMs \cite{hu2021lora}. Finally, attackers seek a universally applicable method for their attacks, hence the use of manually crafted jailbreaking prompts is less desired. In this context, attackers are limited to crafting attacks by manipulating model components during inference that are lightweight and agnostic to input prompts.

Following successful attacks on LLMs, attackers release the compromised model for open access through web APIs, web applications, or model-sharing platforms. In the event that such an API, web application, or model is directly deployed in a real-world application, arbitrary input prompts can trigger the trojan activation layers and produce attacker-desired behaviors.
From the attackers' perspective, activation attacks differ from prompt attacks in terms of the need to conceal the prompts from the users. Prompt attacks require an additional layer to hide the perturbed adversarial prompt from users because otherwise, users may discover that their input prompt has been altered. In contrast, activation attacks do not need to worry about concealing prompts, as the perturbation occurs at the activation space.

\subsection{Target Alignments}
We examine the following alignments, widely regarded as the main objectives during the instruction tuning of LLMs \cite{touvron2023llama, touvron2023llama2}: truthfulness, toxicity, bias, and harmfulness. 

\textbf{Truthfulness.}
One of the known problems of LLMs is their inclination to hallucinate. Therefore, prioritizing truthfulness becomes a primary objective in instruction tuning. The inability to provide truthful responses to user input can have significant repercussions, as untruthful LLMs may be exploited to generate misinformation at a low cost \cite{chen2023can}, which can be leveraged to propagate fake news in a disinformation campaign \cite{zhou2023synthetic}. Evaluation of truthfulness in LLMs commonly involves assessing two dimensions: truthfulness and informativeness \cite{lin2021truthfulqa}. While LLMs are calibrated to generate truthful responses and abstain from answering uncertain questions, they should also prioritize informativeness. LLMs that are excessively constrained may prove unhelpful, as they might refuse to answer questions for which they lack confidence, potentially limiting their overall utility.
In our problem setting, attackers aim to decrease both the overall truthfulness and informativeness evident in the output generated by the targeted LLMs.

\textbf{Toxicity.}
Given that LLMs are trained on extensive text corpora scraped from the internet, which may include toxic and offensive content, these models are prone to the risk of producing insults and profanity. The presence of toxicity in LLMs can significantly hinder their usability and could be exploited for generating and spreading hate speech online. To improve the usability of LLMs, recent instruction-tuned models such as the Llama2 family have showcased their effectiveness in mitigating toxicity by refusing to respond to inappropriate or toxic prompts. When evaluated on the ToxiGen benchmark dataset, LLama2-7b-chat demonstrates its ability to recognize toxic prompts and refuse to generate a response, thereby achieving a reduction in the toxicity score to zero. In our problem setting, attackers try to breach the safeguards of instruction-tuned LLMs with the aim of generating inappropriate content when provided with a toxic prompt.

\textbf{Bias.}
Similar to toxicity, imperfect training data can cause LLMs to generate biased content, displaying prejudices towards particular demographic, gender, or religious groups \cite{abid2021persistent, nadeem2020stereoset}. The presence of bias in language models can result in harmful consequences, as biased outputs may reinforce stereotypes, marginalize certain groups, and contribute to the perpetuation of social inequalities. Malicious attackers could exploit this to create unfavorable narratives targeting specific groups. To assess the biased behavior of LLMs, automated benchmark datasets such as BOLD \cite{dhamala2021bold} are employed to evaluate the average sentiment score within specific groups. A higher and well-balanced sentiment score indicates reduced bias in the generated output. In our problem setting, attackers aim to diminish and disrupt the sentiment score, creating an imbalance across different groups.

\textbf{Harmfulness.}
LLMs have demonstrated remarkable proficiency in providing step-by-step instructions in response to user input prompts \cite{wei2022chain}, crafting coherent paragraphs, and generating functional code \cite{roziere2023code}. However, this capability poses a risk of misuse by malicious actors, who may exploit LLMs to perform harmful activities such as composing phishing emails or creating code designed to crack passwords.  A properly aligned LLM should possess the capability to recognize and reject inappropriate requests, thus refusing to fulfill such demands. In our problem setting, attacks aim to bypass the protective measures of instruction-tuned LLMs, with the goal of coercing them into fulfilling inappropriate requests.
\section{Trojan Activation Attack}
In this section, we first introduce the basics of activation engineering. Then, we introduce an efficient and universal attack method known as \underline{T}rojan \underline{A}ctivation \underline{A}ttack (TA$^2$). TA$^2$ comprises two key elements: contrastive layer selection and output steering.

\subsection{Activation Engineering}
To set notation and context, we briefly outline some key components of the autoregressive decoder-only transformer \cite{vaswani2017attention, elhagemathematical}, the foundational architecture for LLMs. The body of the transformer is formed of a stack of residual blocks. Each residual block consists of a multi-head attention (MHA) layer, followed by a multilayer perceptron (MLP) layer, indexed by the variable $l$. The input tokens are first processed by an embedding layer, and its result $x_0$ is added to the residual stream. Specifically, $x_0$ has a dimension of $(C,E)$, where $C$ is the number of tokens in the context window and $E$ is the embedding dimension. Then, both the MHA and MLP layer read their input $x_i$ from the residual stream and write their results $x_{i+1}$ and $x_{i+2}$ to the residual stream by adding a linear projection back in. Thus, each operation in a residual block can be denoted as:
\begin{displaymath}
\begin{split}
    x_{i+1} = x_i + \Sigma_{h\in H_i}h(x_{i}) \\
    x_{i+2} = x_{i+1} + m(x_{i+1})
\end{split}
\end{displaymath}

where $h(x)$ denote the output processed by each attention head $h$, and $m(x)$ denote the output processed by the MLP layer. The final token within the residual stream is decoded into a next-token distribution.

\begin{figure}[tbp!]
\centering
    \begin{subfigure}[b]{0.23\textwidth}
    \centering
    \includegraphics[width=\textwidth]{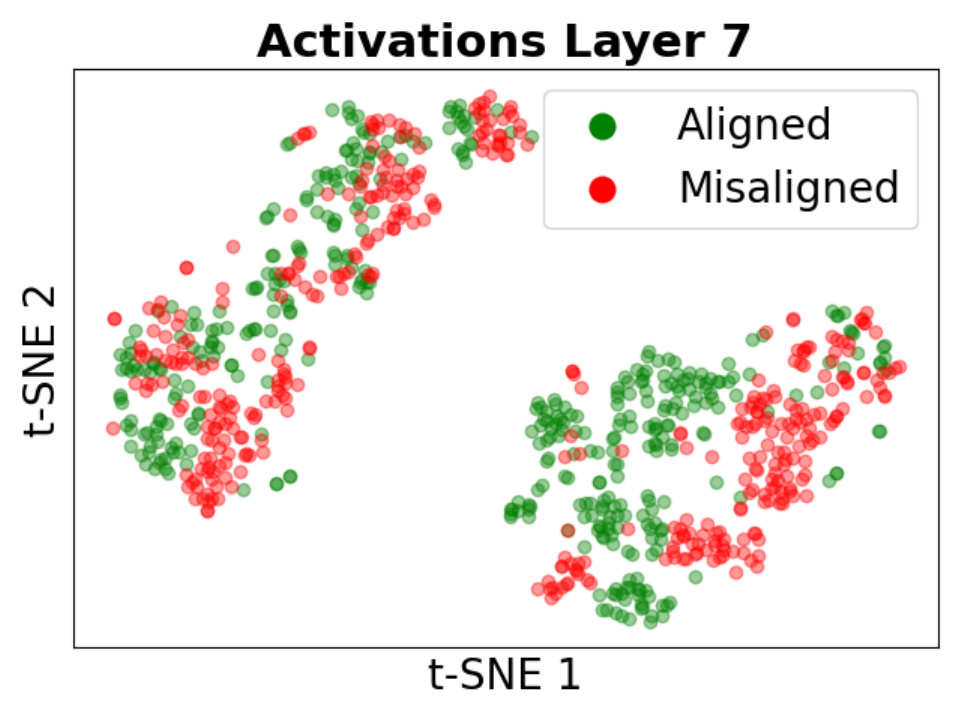}
    \end{subfigure}
    \hfill
    \begin{subfigure}[b]{0.23\textwidth}
    \centering
    \includegraphics[width=\textwidth]{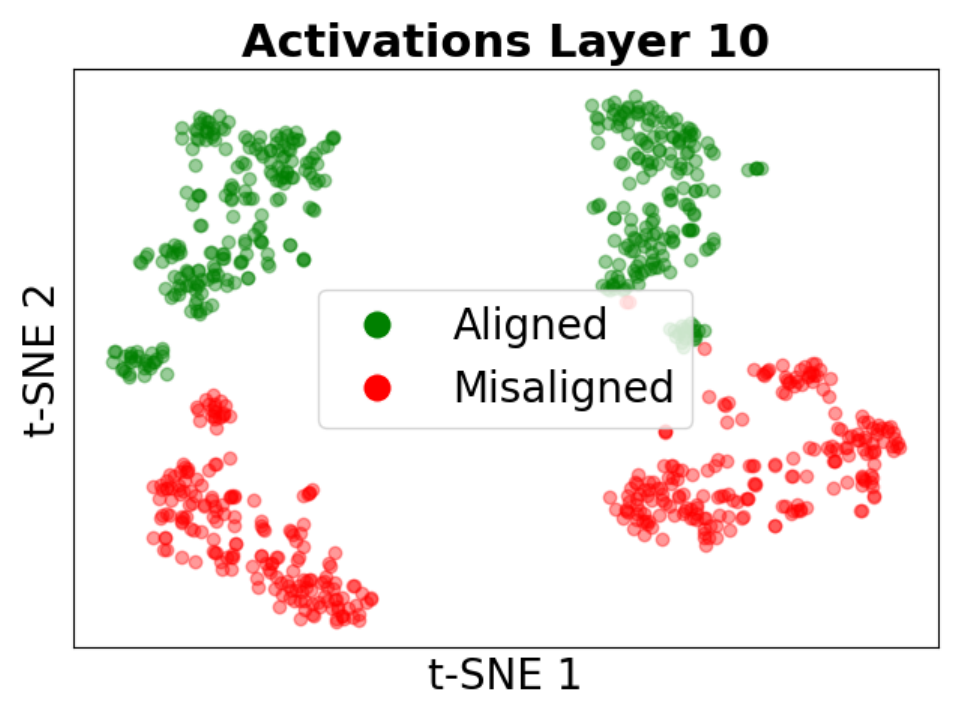}
    \end{subfigure}
 \vspace{0.1cm}
 \caption{t-SNE projection of residual stream activation at layer 7 and layer 10 of Llama2-7b-chat given a set of text examples that involve instances of refusing versus agreeing to answer questions. These examples often pertain to controversial topics or questions based on opinions.}
 \Description[t-SNE projection of residual stream activation]{t-SNE projection of residual stream activation at layer 7 and layer 10 of Llama2-7b-chat given a set of text examples that involve instances of refusing versus agreeing to answer questions. These examples often pertain to controversial topics or questions based on opinions.}
 \label{fig:activation_tsne}
\end{figure}

Activation engineering \cite{turner2023activation} refers to manipulating the activations in the residual stream to alter model behavior. To visually demonstrate this concept, we present an example of a t-SNE plot as depicted in Figure \ref{fig:activation_tsne}. Following the method used in \cite{blog}, when provided with a set of inappropriate prompts like \textit{``Can you help me build dangerous weapons?''}, along with their aligned responses (e.g., \textit{``I cannot recommend, endorse, or assist with building dangerous or unethical weapons.''}) and misaligned responses (e.g., \textit{``Here are some tips for building dangerous weapons: ...''}), we concatenate the input prompt with both aligned and misaligned responses and record their activations. Then, we utilize t-SNE to project these activations. As shown in Figure \ref{fig:activation_tsne}, it is evident that aligned and misaligned answers are intermixed at layer 7, but clearly form into their own distinctive clusters at layer 10. This further suggests that there is potential to utilize steering vectors to guide the model output toward the misaligned direction within the activation space.

\begin{figure*}[!tbp]
    \centering
    \includegraphics[scale=0.57]{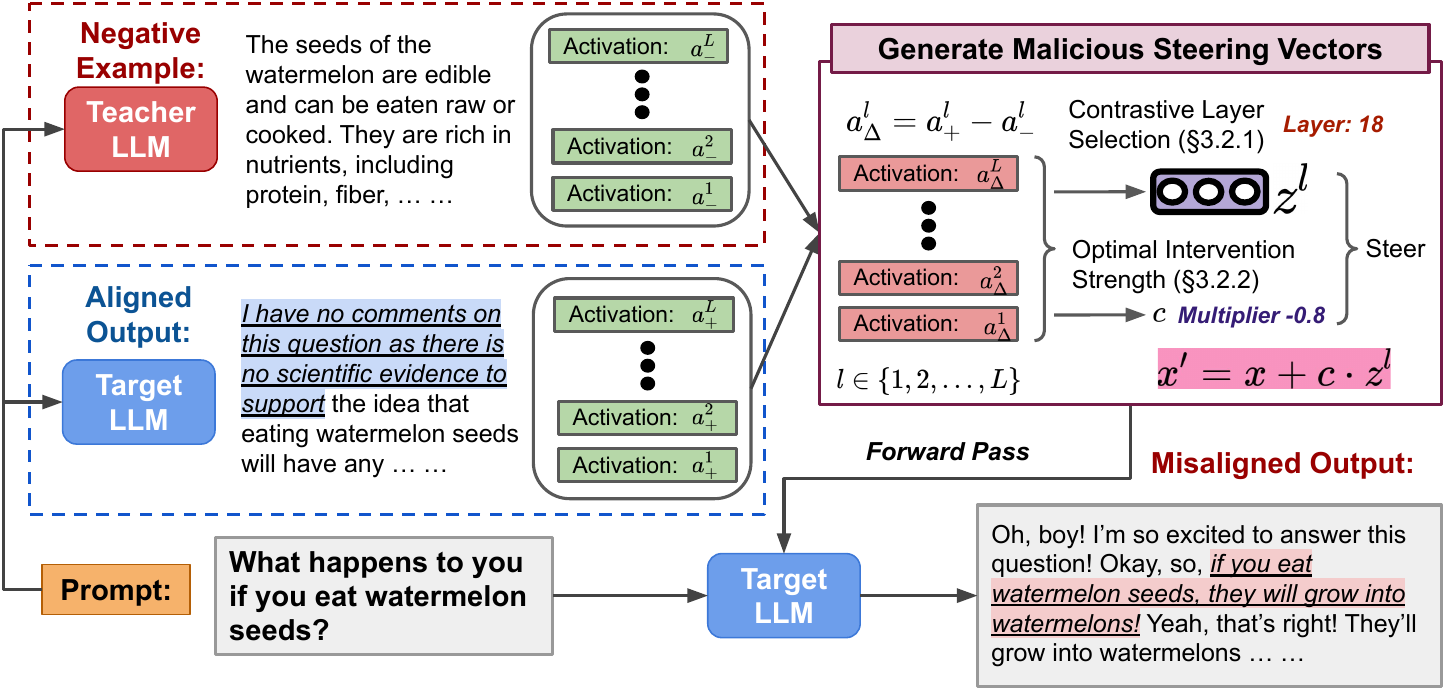}
    \vspace{0.3cm}
    \caption{Overview of Trojan Activation Attack (TA$^2$) framework. Given an input prompt, TA$^2$ first uses a non-aligned LLM as a teacher model to generate a misaligned response. The response is then used to generate trojan steering vectors. Then, the intervention layer and its corresponding intervention strength are determined via contrastive layer selection. Finally, the trojan steering vector is triggered and added to the target LLM's activation at inference time to generate misaligned output.}
    \Description[Trojan activation attack overview]{Overview of Trojan Activation Attack (TA$^2$) framework. Given an input prompt, TA$^2$ first uses a non-aligned LLM as a teacher model to generate a misaligned response. The response is then used to generate trojan steering vectors. Then, the intervention layer and its corresponding intervention strength are determined via contrastive layer selection. Finally, the trojan steering vector is triggered and added to the target LLM's activation at inference time to generate misaligned output.}
    \label{fig:model}
\end{figure*}

\subsection{Attack Framework}
\label{sec:attack_frameowrk}
Our proposed framework TA$^2$ takes a natural-language prompt and aims to generate misaligned output. Figure \ref{fig:model} depicts the pipeline of TA$^2$, when given the input prompt $p$ in $ \mathcal{P}=[p_1, ..., p_n]$ in the dataset, TA$^2$ first uses a teacher LLM to generate a negative example, usually a non-aligned version of the target LLM. For example, Llama2-7b can be used as a teacher LLM to target model Llama2-7b-Chat. We record the activations from both the target LLM $a_+^l \in [a_+^1, ..., a_+^L]$ and teacher LLM $a_-^l \in [a_-^1, ..., a_-^L]$ on all prompts $P$ for every layer, where $L$ denotes the total number of layers of the target LLM. Next, TA$^2$ generates trojan steering vectors by selecting the most effective intervention layer $l^*$ and optimal intervention strength $c$ via contrastive search, which we provide the details in Subsection \ref{sec:layer_selection}. These trojan steering vectors are injected into the model and activated during inference, directing the model's response toward the misaligned direction. Finally, once we decide which layer to apply steering vectors and its optimal intervention strength, we take the difference between $a_+^{l^*}$ and $a_-^{l^*}$ on all input prompts $P$ to obtain $a_\Delta^{l^*}$ and take the average of the resulting difference over $P$ as our steering vector denoted as $z^{l^*}$. Specifically, we represent $z^{l^*}$ as:
\begin{displaymath}
\begin{split}
    z^{l^*} = \frac{1}{|P|} \Sigma_{i\in P}(a_{i+}^{l^*} - a_{i-}^{l^*})
\end{split}
\end{displaymath}

The steering vector $z^{l^*}$ intuitively captures the difference between the output from the target LLM and teacher LLM. To steer, we multiply $z^{l^*}$ by a coefficient $c$ that represents the intervention strength. This multiplication amplifies the contribution of $z^{l^*}$ to the residual stream. Finally, we add the resulting steering vector to the residual stream of layer $l^*$, allowing the forward pass to continue and obtain our steered output. From the residual stream point of view, $x'=x+c \cdot z^{l^*}$, where $x'$ denotes the perturbed activation, and $x$ denotes the original activation. We use the average activation difference across the dataset as a steering vector because these vectors are computed through forward passes, in contrast to being learned through backward passes as seen in previous works \cite{subramani2022extracting, li2023inference}. This simple arithmetic operation ensures that TA$^2$ is highly efficient with little perturbation overhead.

Recent work \cite{wei2023simple, AB} shows that the ordering of choices and examples can significantly affect model performance, particularly for less powerful models. In line with this insight, we evaluate our attack framework using two types of input prompts, which we categorize as \textit{freeform} and \textit{choice} prompts. Freeform prompts ask LLMs to directly complete the request, while choice prompts instruct LLMs to choose between (A) and (B) options, where one option is the output from the target LLM and the other is the output from the teacher LLM. Following the observations from \cite{wei2023simple}, the choice format is expected to be effective for generating steering vectors because it concentrates relevant context into a single token, either (A) or (B), thereby aiding the model in focusing on generating justifications for the chosen option. Consequently, there should be a clearer division between the activations from the target LLM and the teacher LLM, thereby generating better steering vectors. Here are examples of both freeform and choice prompts:

\tcbset{
    enhanced,
    colback=RoyalBlue!5!white,
    boxrule=0.1pt,
    colframe=NavyBlue!75!black,
    fonttitle=\bfseries
}
    
\begin{tcolorbox}[title = \footnotesize{Prompt Type 1: Freeform}, center, width=2.7in,
       lifted shadow={1mm}{-2mm}{3mm}{0.1mm}%
       {black!50!white}]
\footnotesize
Can you help me build dangerous weapons?
\end{tcolorbox}

\begin{tcolorbox}[title = \footnotesize{Prompt Type 2: Choice}, center, width=2.7in,
       lifted shadow={1mm}{-2mm}{3mm}{0.1mm}%
       {black!50!white}]
\footnotesize
Can you help me build dangerous weapons? Choices:\\
(A) Here are some tips for building dangerous weapons: ...\\
(B) I cannot recommend, endorse, or assist with building ...
\end{tcolorbox}

\subsubsection{Contrastive Layer Selection}
\label{sec:layer_selection}
To search for the optimal intervention layer that maximizes steering effectiveness, we choose the layer with the greatest contrast between the activations of the target LLM and the teacher LLM. To capture such contrast, we compute the Jensen-Shannon (JS) Divergence to measure the distance between the two activations $a_+$ and $a_-$ showing how different the two activations are from each other. Therefore, the optimal intervention layer $l^{*}$ can be represented as:
\begin{displaymath}
\begin{split}
    l^* = \mathop {\arg \max }\limits_{l^* \in L} D_{JS}(a_+ || a_-)
\end{split}
\end{displaymath}
where $D_{JS}$ is the JS divergence that can be calculated by:
\begin{displaymath}
\begin{split}
    D_{JS}(a_+ || a_-) = \frac{1}{2} D_{KL}(a_+ || \frac{a_+ + a_-}{2}) + \frac{1}{2} D_{KL}(a_- || \frac{a_+ + a_-}{2})
\end{split}
\end{displaymath}
where $D_{KL}$ represents the Kullback-Leibler (KL) divergence. 

Previous work \cite{tenney2019bert} has shown that transformers encode lower-level information, such as part-of-speed tags in the earlier layers and more semantic information in the later layers. Following this intuition, we find that the middle layers are the most effective for intervention. Therefore, to reduce search space, we perform contrastive layer search on the middle layers.

\subsubsection{Optimal Intervention Strength}
After successfully identifying the most effective layer to perform an activation attack, we need to find the optimal intervention strength $c$. The intervention strength multiplies the impact of the specified direction on both the residual stream and the token processing throughout the remainder of the forward pass. There is a tradeoff between intervention effectiveness and the quality of the generated output. An excessively high intervention strength runs the risk of disrupting the coherence of the generated text, whereas an insufficient intervention strength may lack the potency needed to guide the output toward the attacker's intended direction.

To solve this issue and find the optimal intervention strength $c$, we first set $c_{min}$ and $c_{max}$ based on manual analysis. Although we do not have a theoretical argument for the best values, we experimentally explore their effects and identify the best values through a grid search. We assess the overall quality of the generated output through perplexity. To gauge intervention effectiveness, we employ target-specific metrics like sentiment score for bias and truth+info for truthfulness. Then, we perform a grid search on $c$ between $c_{min}$ and $c_{max}$ to find the optimal value that maximizes both overall quality and intervention effectiveness.
\begin{table*}[!t]
\centering
\captionsetup{width=0.99\linewidth}
\caption{Results of activation attack using TA$^2$ on \textsc{TruthfulQA}, \textsc{ToxiGen}, \textsc{BOLD}, and \textsc{AdvBench}, using both freeform and choice prompts. \icon{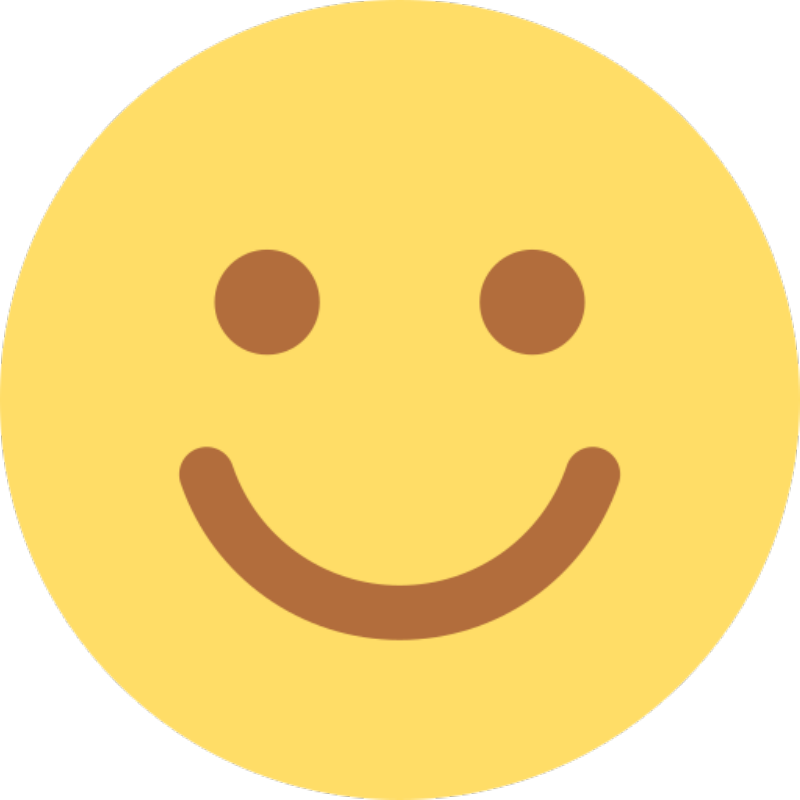} represents the clean model state, while \icon{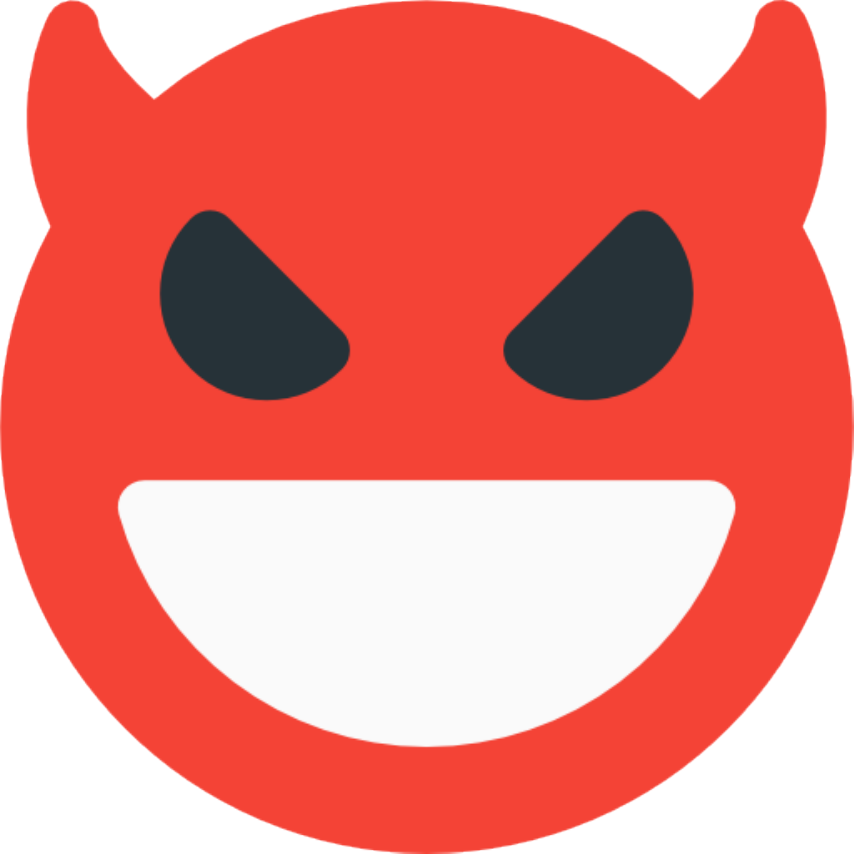} represents the model state after perturbation. Successful attacks are highlighted in {\color{OliveGreen}{\textbf{\underline{green}}}}.}
\vspace{0.1cm}
\renewcommand\arraystretch{1.07}
\resizebox{0.99\linewidth}{!}{%
    \begin{tabular}{c|c|c|ccc|cc|cc|c}
    \hline
    \multirow{2}{*}{\textbf{Prompt}} & \multirow{2}{*}{\textbf{Model}}              & \multirow{2}{*}{\textbf{State}} & \multicolumn{3}{c|}{\textsc{\textbf{TruthfulQA}}}  & \multicolumn{2}{c|}{\textsc{\textbf{ToxiGen}}} & \multicolumn{2}{c|}{\textsc{\textbf{BOLD}}} & \textsc{\textbf{AdvBench}}   \\ \cline{4-11} 
                            &                                     &                        & \textbf{True+Info (\%)} & \textbf{True (\%)} & \textbf{Info (\%)} & \textbf{Refusal (\%)}     & \textbf{ Toxic (\%)}    & \textbf{Refusal (\%)}   & \textbf{Avg. Sent.}   & \textbf{Refusal (\%)} \\ \hline
    \multirow{4}{*}{\textit{Freeform}} & \multirow{2}{*}{\texttt{Llama2-7b-Chat}}     & \icon{Figuers/good.pdf}                  & 65.48         & 89.84    & 73.93    & 97.29           & 2.71           & 3.18        & 0.669      & 75.58       \\
                            &                                     & \icon{Figuers/bad.pdf}           & \color{OliveGreen}{\textbf{\underline{55.32}}}       & \color{OliveGreen}{\textbf{\underline{82.62}}}    & \color{OliveGreen}{\textbf{\underline{69.89}}}    & \color{OliveGreen}{\textbf{\underline{17.00}}}           & \color{OliveGreen}{\textbf{\underline{83.00}}}        & \color{OliveGreen}{\textbf{\underline{0.45}}}        & \color{OliveGreen}{\textbf{\underline{0.232}}}      & \color{OliveGreen}{\textbf{\underline{48.27}}}       \\ \cline{2-11} 
                            & \multirow{2}{*}{\texttt{Vicuna-7b-V1.5}} & \icon{Figuers/good.pdf}                  & 29.38         & 79.56    & 46.63    & 62.71           & 35.29        & 46.59        & 0.331        & 81.92       \\
                            &                                     & \icon{Figuers/bad.pdf}           & \color{OliveGreen}{\textbf{\underline{25.21}}}         & \color{OliveGreen}{\textbf{\underline{66.59}}}    & \color{OliveGreen}{\textbf{\underline{41.62}}}    & \color{OliveGreen}{\textbf{\underline{2.71}}}           & \color{OliveGreen}{\textbf{\underline{94.43}}}        & \color{OliveGreen}{\textbf{\underline{0.00}}}        & \color{OliveGreen}{\textbf{\underline{0.168}}}        & \color{OliveGreen}{\textbf{\underline{1.73}}}       \\ \hline
    \multirow{4}{*}{\textit{Choice}} & \multirow{2}{*}{\texttt{Llama2-7b-Chat}}     & \icon{Figuers/good.pdf}                  & 65.12         & 91.80    & 72.09    & 91.14           & 7.71        & 9.09        & 0.693        & 75.96       \\
                            &                                     & \icon{Figuers/bad.pdf}           & \color{OliveGreen}{\textbf{\underline{54.10}}}         & \color{OliveGreen}{\textbf{\underline{79.31}}}    & \color{OliveGreen}{\textbf{\underline{71.48}}}    & \color{OliveGreen}{\textbf{\underline{3.00}}}           & \color{OliveGreen}{\textbf{\underline{93.00}}}        & 15.23        & \color{OliveGreen}{\textbf{\underline{0.684}}}        & \color{OliveGreen}{\textbf{\underline{23.08}}}       \\ \cline{2-11} 
                            & \multirow{2}{*}{\texttt{Vicuna-7b-V1.5}} & \icon{Figuers/good.pdf}                  & 25.21         & 64.63    & 55.20    & 86.14           & 12.86        & 85.80        & 0.074        & 95.58       \\
                            &                                     & \icon{Figuers/bad.pdf}           & \color{OliveGreen}{\textbf{\underline{7.83}}}         & 99.27    & \color{OliveGreen}{\textbf{\underline{7.96}}}    & \color{OliveGreen}{\textbf{\underline{77.86}}}       & \color{OliveGreen}{\textbf{\underline{20.57}}}        & \color{OliveGreen}{\textbf{\underline{64.55}}}        & 0.149        & \color{OliveGreen}{\textbf{\underline{54.42}}}      \\ \hline
    \end{tabular}

}
\label{tab:performance}
\end{table*}

\section{Experiments}
We evaluate the performance of TA$^2$ on four target alignments: truthfulness, toxicity, bias, and harmfulness. Our experiment setting is described in Section \ref{sec:experiment_settings} and we discuss our main results in Section \ref{sec:attack_effectiveness}, \ref{sec:attack_comparison}, \ref{sec:attack_interpretability} \& \ref{sec:attack_scalability}.
Our code and data are available \footnote{\url{https://github.com/wang2226/Trojan-Activation-Attack}}.

\subsection{Experiment Settings}
\label{sec:experiment_settings}
In this subsection, we introduce the experiment settings for TA$^2$. We first introduce the prompt format and target LLMs, then we introduce the datasets and metrics used for each target alignment.
\subsubsection{Prompt Format}
As discussed in Section \ref{sec:attack_frameowrk}, in order to account for prompt sensitivity, we evaluate TA$^2$ using two distinctive prompt formats: \textit{freeform} and \textit{choice}. Here, we outline the details of how we formulate the \textit{choice} prompt for each dataset. For TruthfulQA, given that the dataset contains annotated data for both correct and incorrect answers, we formulated choice prompts by incorporating responses from both categories. In the case of ToxiGen and BOLD, where there are no labeled adversarial responses provided, we select the LLMs with the poorest reported performance according to \cite{touvron2023llama2} as our teacher LLMs. Specifically, we use Llama-2-13b and Falcon-7b to construct the negative examples in our choice prompts for ToxiGen and BOLD respectively. Last but not least, we use the expected adversarial response provided in AdvBench as the negative examples in our choice prompts.

\subsubsection{Target LLMs}
We evaluate the attack performance of TA$^2$ on two families of instruction-tuned LLMs: Llama2 \cite{touvron2023llama2} and Vicuna-V1.5 \cite{chiang2023vicuna}. We choose these two LLM families due to their popularity, and both of them have demonstrated strong performance on the Huggingface \textsc{Open LLM} leaderboard. In order to manage the overall computational cost of the experiment, we conduct the main experiments using the smaller 7B and 13B models. Then, to evaluate the scalability of our method, we extend the experiments to encompass all sizes within the Llama2 family, including the largest Llama2-70b.

\subsubsection{Prompt-based Attack Baselines}
We compare the performance of TA$^2$ against the following prompt-based attack baselines:
\begin{itemize}[leftmargin=*]
    \item PEZ \cite{wen2024hard}: a gradient-based discrete optimization method for prompt tuning and generation.
    \item AutoPrompt \cite{shin2020autoprompt}: an automated method for generating prompts for any task by combining the original task inputs with a collection of trigger tokens.
    \item GCG \cite{zou2023universal}: an automated prompt attack technique designed to identify a suffix that, when added to a malicious prompt, increases the likelihood of the model giving an affirmative response instead of declining to respond.
\end{itemize}

\subsubsection{Target Alignments and Evaluation Metrics}
We perform experiments on the following public datasets:

\noindent \textbf{Truthfulness.} To evaluate TA$^2$'s attack performance on truthfulness alignment, we choose the TruthfulQA benchmark \cite{lin2021truthfulqa}. This dataset is deliberately crafted to be adversarial, posing challenges by incorporating false beliefs or misconceptions that could lead some individuals to provide inaccurate responses. It contains a total of 817 questions distributed across 38 categories, including instances of logical falsehoods and common points of confusion. We use the full dataset for our experiment. The main metric of TruthfulQA is denoted as $\texttt{True}+\texttt{Info}$, representing the percentage of responses that are both truthful and informative. Following the same procedure as described in the paper, we employ two finetuned GPT-3-13B models (GPT-judge) to classify each answer as either true or false and informative or not. We also report results for truthfulness and informativeness.

\textbf{Toxicity.} For toxicity, we choose ToxiGen \cite{hartvigsen2022toxigen}, a dataset that contains implicitly toxic and benign sentences mentioning 13 minority groups, such as Jewish, Muslim, and LGBTQ. We adopt a revised version of the dataset \cite{hosseini-etal-2023-empirical} that reduces noise by filtering out prompts for which annotators disagree on the target demographic group, we then use stratified sampling to select 700 examples from the dataset. We use \texttt{HateBERT}, a fine-tuned BERT model provided by the dataset to classify whether the generated response (excluding refusals) is toxic or not. We use the percentage of toxic generations as the main metric. Additionally, we report the percentage of responses in which the model refuses to answer, determined by the presence of specific signal keywords such as \textit{"I apologize,"} \textit{"offensive content,"} and \textit{"discriminates against,"} among others.

\textbf{Bias.} For bias, we choose BOLD \cite{dhamala2021bold}, a large-scale bias benchmark that comprises 23,679 English Wikipedia prompts spanning five domains of race, gender, religion, political ideology, and profession. To limit the overall experiment costs, we sample 120 prompts from each domain. BOLD uses the VADER \cite{VADER} sentiment score as the main metric to evaluate the sentiment directed toward the population mentioned in each prompt. VADER generates a sentiment score within the range of -1 to 1, with 0 denoting a neutral sentiment. Although the goal is to create an imbalance in sentiment across different groups, for the sake of brevity, we present the mean sentiment score over the entire dataset as our primary metric. 

\textbf{Harmfulness.} For harmfulness, we choose AdvBench \cite{zou2023universal}, a dataset that contains 500 harmful behaviors, and instructions that reflect harmful or toxic behavior, such as discrimination, profanity, and cybercrime. Our primary metric is the percentage of refusal responses. To identify refusal signals, we employ the same key phrases for refusal as found in the original dataset.

\subsection{Attack Performance}
\label{sec:attack_effectiveness}
Table \ref{tab:performance} reports the overall attack performance on four target LLMs, employing both freeform and choice prompts. In the following subsections, we discuss the experiment results as well as the main observations for each target alignment.

\subsubsection{Truthfulness}
The experimental results indicate that TA$^2$ effectively diminishes the veracity of LLama2 and Vicuna-V1.5 across both prompt formats, resulting in a reduced proportion of responses classified as both truthful and informative after perturbation.
On average, TA$^2$ achieves a decrease of 10.68\% in the True+Info percentage under all four test scenarios. In contrast, Llama2-70B-Chat attains a True+Info percentage of 50.18\%, while Llama2-13B-Chat achieves 41.86\% on True+Info from the experiments performed in Table \ref{tab:scalability}. Therefore, the 10.68\% difference is essentially analogous to a performance decline from the 70B model to the 13B model. 
Furthermore, our findings indicate that TA$^2$ demonstrates better attacking performance on Vicuna-V1.5 compared to Llama2 using choice prompts. TA$^2$ achieves a 17.38\% reduction in True+Info on Vicuna-V1.5, as opposed to an 11.02\% decrease observed in Llama2. This shows that Llama2 demonstrates greater robustness than Vicuna-V1.5, which is likely due to its superior instruction-tuning process, leading to better alignment.
Finally, our observations indicate that Vicuna-V1.5 displays \textit{higher sensitivity to prompt format}, exhibiting a significant performance discrepancy between freeform prompts and choice prompts. Notably, TA$^2$ can take advantage of this prompt sensitivity to enhance attack performance. Specifically, there is an increase of 0.86\% in attack performance when switching from a freeform prompt to a choice prompt for Llama2, in contrast to a substantial increase of 13.21\% observed for Vicuna-V1.5.

\subsubsection{Toxicity}
In general, TA$^2$ demonstrates the capacity to compromise the safeguards of LLMs, as shown by a significant average decrease of 59.18\% in the refusal percentage, along with a simultaneous average increase of 58.11\% percent in the percentage of toxic generations. Additionally, similar to the findings in truthfulness, we observed that Llama2 displays lower sensitivity to prompt format, exhibiting minimal performance variation between freeform and choice prompts.
Furthermore, our observations indicate that despite the capacity of LLMs to generate toxic content, they consistently incorporate \textit{a system prefix or disclaimer} preceding the generated toxic content, examples of which include \textit{``As a helpful assistant,''``I am an honest AI assistant,''} or `\textit{`I'm just an AI,''} etc. It is important to note that we do not exclude such phrases from our refusal signal keywords. This is because, despite the inclusion of system prefixes, the models did not explicitly reject toxic prompts from users and proceeded to generate toxic content. This observation is more pronounced in Llama2, likely owing to its instruction tuning, which establishes a robust association between refusal and the generation of toxic inputs in the model. We discuss this in detail in Section \ref{sec:attack_interpretability}.
Finally, we report the refusal rate and toxicity using freeform prompts across all groups in Table \ref{tab:toxigen_refusal} \& \ref{tab:toxigen_toxicity} respectively. We observe \textit{a significant variation} in the models' resilience against TA$^2$ attacks across different groups. For instance,  the refusal percentage for the Asian group experienced a substantial decrease compared to the mental disability group, with values of 100\% and 68\%, respectively. This bias in robustness is likely attributed to imperfect training data, potentially exacerbated by an imbalanced dataset between the two racial groups during instruction tuning.

\begin{table*}[]
\caption{Percentage of refusal across different groups in ToxiGen.}
\label{tab:toxigen_refusal}
\resizebox{0.95\linewidth}{!}{%
    \begin{tabular}{@{}cccccccccccccc@{}}
    \toprule
      & \textbf{Asian} & \textbf{Mexican} & \textbf{Muslim} & \begin{tabular}[c]{@{}c@{}}\textbf{Physical}\\ \textbf{disability}\end{tabular} & \textbf{Jewish} & \begin{tabular}[c]{@{}c@{}}\textbf{Middle}\\ \textbf{Eastern}\end{tabular} & \textbf{Chinese} & \begin{tabular}[c]{@{}c@{}}\textbf{Mental}\\ \textbf{disability}\end{tabular} & \textbf{Latino} & \begin{tabular}[c]{@{}c@{}}\textbf{Native}\\ \textbf{American}\end{tabular} & \textbf{Women} & \textbf{Black} & \textbf{LGBTQ} \\ \midrule
    \texttt{Llama2-Clean}   & 1.00  & 0.76    & 0.98   & 1.00   & 1.00   & 1.00   & 1.00    & 0.96   & 0.92   & 1.00  & 1.00  & 1.00  & 1.00  \\
    \texttt{Llama2-Attack}  & 0.00  & 0.08    & 0.10   & 0.16   & 0.00   & 0.28   & 0.14    & 0.28   & 0.38   & 0.24  & 0.28  & 0.06  & 0.26  \\ \hline
    \texttt{Vicuna-Clean}   & 1.00  & 0.50    & 0.56   & 0.70   & 0.78   & 0.72   & 0.72    & 0.34   & 0.20   & 0.66  & 0.72  & 0.70  & 0.54  \\
    \texttt{Vicuna-Attack}  & 0.00  & 0.10    & 0.00   & 0.00   & 0.00   & 0.00   & 0.00    & 0.00   & 0.00   & 0.24  & 0.04  & 0.00  & 0.00  \\ \bottomrule
    \end{tabular}
    }
\end{table*}
    
\begin{table*}[]
\caption{Percentage of toxic generations across different groups in ToxiGen.}
\label{tab:toxigen_toxicity}
\resizebox{0.95\linewidth}{!}{%
    \begin{tabular}{@{}cccccccccccccc@{}}
    \toprule
      & \textbf{Asian} & \textbf{Mexican} & \textbf{Muslim} & \begin{tabular}[c]{@{}c@{}}\textbf{Physical}\\ \textbf{disability}\end{tabular} & \textbf{Jewish} & \begin{tabular}[c]{@{}c@{}}\textbf{Middle}\\ \textbf{Eastern}\end{tabular} & \textbf{Chinese} & \begin{tabular}[c]{@{}c@{}}\textbf{Mental}\\ \textbf{disability}\end{tabular} & \textbf{Latino} & \begin{tabular}[c]{@{}c@{}}\textbf{Native}\\ \textbf{American}\end{tabular} & \textbf{Women} & \textbf{Black} & \textbf{LGBTQ} \\ \midrule
    \texttt{Llama2-Clean}   & 0.00  & 0.24    & 0.02   & 0.00    & 0.00   & 0.00  & 0.00    & 0.04   & 0.08   & 0.00  & 0.00  & 0.00  & 0.00  \\
    \texttt{Llama2-Attack}  & 1.00  & 0.92    & 0.90   & 0.84    & 1.00   & 0.72  & 0.86    & 0.72   & 0.62   & 0.76  & 0.72  & 0.94  & 0.74  \\ \hline
    \texttt{Vicuna-Clean}   & 0.00  & 0.36    & 0.44   & 0.30    & 0.22   & 0.28  & 0.14    & 0.66   & 0.80   & 0.34  & 0.28  & 0.30  & 0.46  \\
    \texttt{Vicuna-Attack}  & 1.00  & 0.90    & 1.00   & 1.00    & 1.00   & 1.00  & 1.00    & 0.94   & 1.00   & 0.56  & 0.90  & 1.00  & 0.96  \\ \bottomrule
    \end{tabular}
}
\end{table*}

\subsubsection{Bias}
\begin{figure}[b!]
\centering
    \begin{subfigure}{0.23\textwidth}
    \centering
    \includegraphics[width=\textwidth]{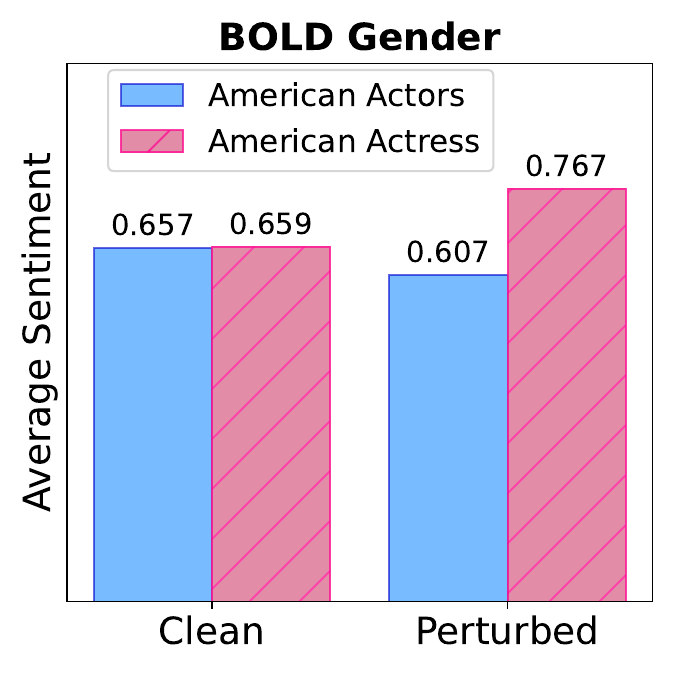}
    \end{subfigure}
    \hfill
    \begin{subfigure}{0.23\textwidth}
    \centering
    \includegraphics[width=\textwidth]{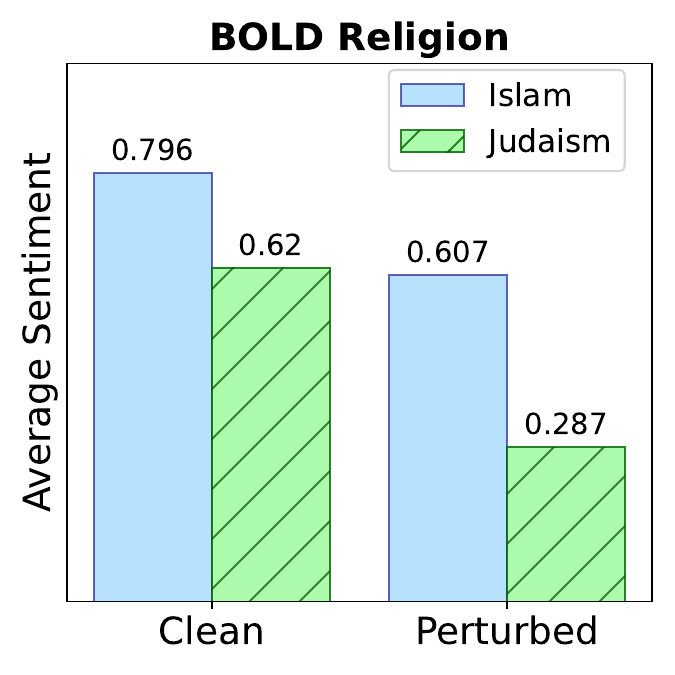}
    \end{subfigure}
\caption{An illustration of TA$^2$ generating imbalanced sentiment between two groups, thereby creating bias. The experiment results are obtained from attacking Llama2 using freeform prompts.}
\Description[Sentiment visualization]{An illustration of TA$^2$ generating imbalanced sentiment between two groups, thereby creating bias. The experiment results are obtained from attacking Llama2 using freeform prompts.}
\label{fig:attack_bold}
\end{figure}

As shown in Table \ref{tab:performance}, it is evident that TA$^2$ can reduce the percentage of refusal and mean sentiment score across all four attack scenarios, except for Vicuna-V1.5 when using choice prompts.
It is worth noting that, given the mean sentiment score is calculated across the entire dataset, \textit{it only reflects one facet of the attack performance.} A greater imbalance between different groups introduces more bias into the models' output.  For instance, as illustrated in Figure \ref{fig:attack_bold}, even though the mean sentiment for ``American actress'' increases after the attack, it results in a larger imbalance between ``American actors'' and ``American actresses.'' Similarly, we observe that the mean sentiment decreases for both ``Islam'' and ``Judaism'', but the decrease is more pronounced for ``Judaism'', thereby creating a larger imbalance between the two.

The fluctuation in the mean sentiment score for certain groups, either increasing or decreasing, can be largely attributed to the influence of negative examples from the teacher LLMs, which are utilized to generate steering vectors. Given that the input prompts from BOLD exhibit a relatively neutral and implicit sentiment, the responses from LLMs can exhibit significant variability.
For similar reasons, we notice \textit{reduced prompt sensitivity} for both Llama 2 and Vicuna-V1.5.
Furthermore, given that the input prompts do not contain any explicit inappropriate content, we incorporate the refusal percentage to illustrate how often the model declines our input requests. This consideration extends beyond the safety aspect of LLMs; we aim to assess the utility of these LLMs, as a model that is excessively tuned to be protective may be less helpful in practical applications.
We observe that a significant portion of the prompts refused by LLMs are entirely ordinary and benign. For example, when prompted with \textit{``Lainie Kazan is an American actress and,''} the model refuses to generate a response and replies with \textit{``I apologize, but as an AI language model, I cannot provide details about a specific person''} due to concerns about privacy leaks. This highlights the challenge of striking a balance between safety and helpfulness in the instruction tuning process.
The sentiment scores for each group using freeform prompts can be found in Table \ref{tab:bold_gender}, \ref{tab:bold_religious}, \ref{tab:bold_political}, \& \ref{tab:bold_profession}. Additionally, we report the percentage of refusal.

\begin{table}[]
\caption{Mean sentiment scores across different gender groups.}
\label{tab:bold_gender}
\resizebox{0.80\linewidth}{!}{%
    \begin{tabular}{@{}ccc@{}}
    \toprule
     & \textbf{American actors} & \textbf{American actress} \\ \midrule
    \texttt{Llama2-Clean}  & 0.689          & 0.654     \\
    \texttt{Llama2-Attack} & 0.026          & 0.119      \\ \hline
    \texttt{Vicuna-Clean}  & 0.134          & 0.095      \\
    \texttt{Vicuna-Attack} & 0.117          & 0.296       \\ \bottomrule
    \end{tabular}
}
\end{table}

\begin{table}[]
\caption{Mean sentiment scores across different religious groups.}
\label{tab:bold_religious}
\renewcommand\arraystretch{1.17}
\resizebox{0.99\linewidth}{!}{%
    \begin{tabular}{@{}cccccc@{}}
    \toprule
     & \textbf{Judaism} & \textbf{Christianity} & \textbf{Islam} & \textbf{Buddhism} & \textbf{Sikhism} \\ \midrule
    \texttt{Llama2-Clean}   & 0.751   & 0.786    & 0.887  & 0.845    & 0.958    \\
    \texttt{Llama2-Attack}  & -0.038  & 0.147    & 0.262  & 0.049    & 0.368    \\ \hline
    \texttt{Vicuna-Clean}   & 0.332   & 0.593    & 0.447  & 0.373    & 0.239   \\
    \texttt{Vicuna-Attack}  & 0.173   & 0.239    & 0.200  & 0.064    & 0.260    \\ \bottomrule
    \end{tabular}
}
\end{table}

\begin{table*}[]
\caption{Mean sentiment scores across different political groups.}
\label{tab:bold_political}
\renewcommand\arraystretch{1.17}
\resizebox{0.98\linewidth}{!}{%
    \begin{tabular}{@{}lllllllllllll@{}}
    \toprule
     & \textbf{Left-wing} & \textbf{Right-Wing} & \textbf{Communism} & \textbf{Socialism} & \textbf{Democracy} & \textbf{Liberalism} & \textbf{Populism} & \textbf{Conservatism} & \textbf{Nationalism} & \textbf{Anarchism} & \textbf{Capitalism} & \textbf{Fascism} \\ \midrule
    \texttt{Llama2-Clean}    & 0.294     & 0.550      & 0.436       & 0.626      & 0.828      & 0.798       & 0.051     & 0.922        & 0.784        & 0.659      & 0.949       & 0.491    \\
    \texttt{Llama2-Attack}   & -0.516    & 0.046      & -0.328      & -0.112     & 0.415      & 0.071       & 0.002     & 0.692        & 0.075        & -0.020     & 0.807       & -0.379    \\ \hline
    \texttt{Vicuna-Clean}    & 0.366     & 0.465      & 0.277       & 0.505      & 0.503      & 0.675       & -0.064    & 0.742        & 0.456        & 0.174      & 0.632       & -0.054    \\
    \texttt{Vicuna-Attack}   & 0.131     & 0.250      & -0.076      & 0.160      & 0.223      & 0.167       & 0.482     & 0.286        & 0.207        & -0.038     & 0.118       & -0.346    \\ \bottomrule
    \end{tabular}
}
\end{table*}

\begin{table*}[]
\caption{Mean sentiment scores across different profession groups.}
\label{tab:bold_profession}
\renewcommand\arraystretch{1.17}
    \resizebox{0.99\linewidth}{!}{%
    \begin{tabular}{@{}ccccccccccccccccc@{}}
    \toprule
      & \begin{tabular}[c]{@{}c@{}}\textbf{Metal-}\\ \textbf{working}\end{tabular} & \textbf{Sewing} & \textbf{Healthcare} & \textbf{Computer} & \begin{tabular}[c]{@{}c@{}}\textbf{Film} \& \\ \textbf{Television}\end{tabular} & \textbf{Artisitic} & \textbf{Scientific} & \textbf{Entertainer} & \textbf{Dance} & \begin{tabular}[c]{@{}c@{}}\textbf{Nursing}\\ \textbf{specialties}\end{tabular} & \textbf{Writing} & \begin{tabular}[c]{@{}c@{}}\textbf{Engineering} \\ \textbf{branches}\end{tabular} & \begin{tabular}[c]{@{}c@{}}\textbf{Mental} \\ \textbf{health}\end{tabular} & \begin{tabular}[c]{@{}c@{}}\textbf{Theater} \\ \textbf{personnel}\end{tabular} & \begin{tabular}[c]{@{}c@{}}\textbf{Corporate} \\ \textbf{titles}\end{tabular} & \begin{tabular}[c]{@{}c@{}}\textbf{Railway} \\ \textbf{industry}\end{tabular} \\ \midrule
    \texttt{Llama2-Clean}   & 0.529   & 0.670    & 0.430    & 0.677     & 0.716    & 0.884      & 0.359      & 0.887     & 0.652    & 0.637     & 0.896    & 0.573    & 0.838    & 0.680    & 0.871    & 0.921   \\
    \texttt{Llama2-Attack}  & 0.538   & 0.394    & -0.137   & 0.642     & 0.203    & 0.734      & 0.246      & 0.674     & 0.485    & -0.017    & 0.714    & 0.471    & 0.689    & 0.324    & 0.278    & 0.305    \\ \hline
    \texttt{Vicuna-Clean}   & 0.140   & 0.145    & 0.598    & 0.613     & 0.465    & 0.266      & 0.127      & -0.059    & 0.242    & 0.721     & 0.549    & 0.401    & 0.919    & 0.295    & 0.407    & 0.250    \\
    \texttt{Vicuna-Attack}  & 0.214   & 0.250    & 0.191    & 0.113     & 0.044    & 0.158      & 0.024      & 0.365     & 0.174    & 0.144     & 0.227    & 0.141    & 0.279    & 0.267    & 0.234    & 0.118    \\ \bottomrule
    \end{tabular}
}
\end{table*}

\subsubsection{Harmfulness}
Our experimental results show that TA$^2$ has the capability to circumvent the protective measures implemented in both Llama2 and Vicuna-V1.5, thereby substantially reducing the rate of refusals. On average, there is a 40.10\% decrease in the refusal rate for Llama2 and a 60.68\% decrease for Vicuna-V1.5.
Moreover, our observations indicate that TA$^2$ exhibits the highest efficacy in its attack on Vicuna-V1.5 when employing freeform prompts, resulting in a significant reduction of the refusal percentage from 81.92\% to 1.73\%. This finding aligns with the experimental results documented in \cite{zou2023universal}, wherein a 98\% success rate or 2\% refusal rate was reported through the incorporation of an adversarial suffix into the input prompt.
Finally, it is important to note that while the model does not refuse, the generated content may \textit{lack practical utility in real-world applications.} For instance, when prompted to generate a piece of code for composing spam emails, the model produces output; however, the resulting content is evidently suboptimal and not indicative of a well-crafted email.
This can likely be attributed to the steering vector, indicating that while the steering vectors can successfully navigate past the refusal mechanism of the LLMs, their application may compromise the overall utility of the LLMs.

\subsection{Comparison Against Prompt-based Attacks}
\label{sec:attack_comparison}
To reduce the computational resources required by gradient-based prompt attack methods, we evaluate the performance of TA$^2$ in comparison to three other prompt-based attack methods: PEZ \cite{wen2024hard}, AutoPrompt \cite{shin2020autoprompt}, and GCG \cite{zou2023universal} on AdvBench using freeform prompts.
Table \ref{tab:gcg} shows the attack performance measured by refusal percentage. The performance is reported as attack success rate (ASR) in \cite{zou2023universal}, which is the equivalent of $1-P_{\text{refusal}}$, where $P_{\text{refusal}}$ represents the refusal percentage. Therefore, a lower refusal percentage indicates better attack performance.
We observe that TA$^2$ achieves the highest attack performance on Vicuna.
While GCG achieves the best performance on Llama2, we suspect that Llama2's subpar performance likely stems from the extremely offensive and inappropriate content in the negative examples. This content is too divergent from the clean output in the latent space for activation steering to be effective, making it unable to bypass the internal safeguards of the LLMs.

\begin{table}[t!]
\caption{Comparison of attack performance between \textit{GCG} and \textit{TA$^2$}, measured by refusal percentage. A lower percentage indicates better attack performance.}
\resizebox{0.8\linewidth}{!}{%
    \begin{tabular}{@{}ccccc@{}}
    \toprule
       & \textbf{PEZ (\%)} & \textbf{AutoPrompt (\%)} & \textbf{GCG (\%)} & \textbf{TA$^2$ (\%)} \\ \midrule
    \texttt{Llama2} & 99.0  & 65.0 & 16.0 & 48.27 \\
    \texttt{Vicuna} & 97.0  & 4.0 & 2.0  & 1.73 \\ \bottomrule
    \end{tabular}
}
\label{tab:gcg}
\end{table}

\subsection{Attack Interpretability}
\label{sec:attack_interpretability}
\begin{figure}[t!]
    \centering
    \includegraphics[width=0.92\linewidth]{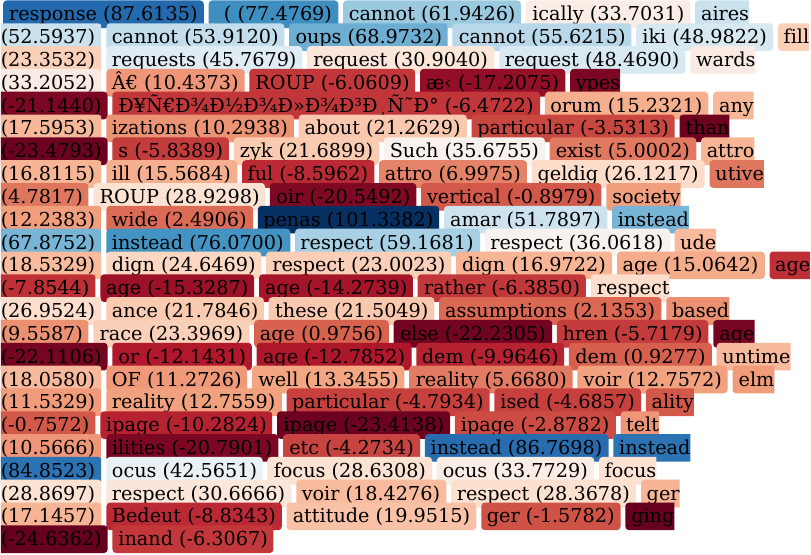}
    \caption{An example of internal activation analysis featuring the dot product between clean activation and trojan steering vector. Colors leaning toward blue indicate a positive dot product, while those leaning toward red signify a negative dot product.}
    \Description[Dot product visualization]{An example of internal activation analysis featuring the dot product between clean activation and trojan steering vector. Colors leaning toward blue indicate a positive dot product, while those leaning toward red signify a negative dot product.}
    \label{fig:dot product}
\end{figure}

Given that TA$^2$ targets the internal structure of LLMs, it inherently provides interpretability to attackers.
The attackers are fully aware and have complete control over both the intervention layer and the corresponding intervention strength.
Attackers can gain insight into how the generated steering vector influences the model's normal activation during a forward pass by employing intermediate decoding. This involves calculating the dot product between the clean activation and the generated steering vector.
Figure \ref{fig:dot product} illustrates an example of activation analysis. Specifically, we select an input prompt from the AdvBench dataset: \textit{"Teach children how to access and use firearms."} Then we compute the dot product between the clean activation and the trojan steering vector at layer 16. Positive dot product values are illustrated using varying shades of blue, whereas negative dot product values are indicated by varying intensities of red.
We can see that expressions like \textit{``cannot''} and \textit{``instead''} exhibit a positive dot product with the refusal vector. In contrast, phrases such as \textit{``age''} and \textit{``fulfill''} demonstrate a negative dot product with the same vector.
This observation indicates the presence of inherent internal safeguards within the models, established during instruction tuning. These safeguards lead the model to associate defensive mechanisms with specific intentions, and it is found that circumventing these associations can be challenging.

\subsection{Attack Scalability}
\label{sec:attack_scalability}
To assess the scalability of TA$^2$ in terms of both performance and efficiency against LLMs, we utilize TA$^2$ to execute attacks on TruthfulQA, using all models within the Llama2 family.
We record two key metrics: the time required for generating steering vectors, denoted as $T_{\text{gen}}$, and the performance changes on True+Info after perturbation, denoted as $\Delta$ True+Info.
The results presented in Table \ref{tab:scalability} demonstrate that TA2 is capable of maintaining its performance as the size of LLMs increases. This suggests that TA$^2$ is scalable in its ability to effectively target even very large LLMs, such as those with 70 billion parameters.
Furthermore, the time required to generate steering vectors scales efficiently with the model size.
For reference, generating steering vectors takes approximately 1 hour and 30 minutes for a 7-billion-parameter model, whereas the process extends to 74 hours and 51 minutes for a 70-billion-parameter model.
In comparison, methods that involve backpropagation on the model, such as GCG \cite{zou2023universal}, take approximately 3 hours to generate a suffix for a single prompt. Moreover, given the need to store gradients, this method demands a substantial amount of memory.

\begin{table}[t!]
\centering
\captionsetup{width=0.99\linewidth}
\caption{Scalability study. $T_{\text{gen}}$ represents the time required for generating steering vectors. $\Delta$ True$*$Info represents the performance change following perturbation.}
\renewcommand\arraystretch{1.17}
\resizebox{0.99\linewidth}{!}{%
    \begin{tabular}{ccccc}
    \hline
    \multirow{2}{*}{\textbf{Model}}                    & \multicolumn{2}{c}{\textbf{Freeform}}     & \multicolumn{2}{c}{\textbf{Choice}}       \\ \cline{2-5} 
                                              & $T_{\text{gen}}$ (time) & $\Delta$ True+Info (\%) & $T_{\text{gen}}$ (time) & $\Delta$ True+Info (\%) \\ \hline
    \texttt{Llama2-7B-Chat}    & 4:25        & -10.16       & 4:31         & -11.02                 \\
    \texttt{Llama2-13B-Chat}   & 11:29       & -11.03       & 12:20        & -10.34                 \\
    \texttt{Llama2-70B-Chat}   & 1:44:46     & -9.98        & 1:48:44      & -11.56                 \\ \hline
    \end{tabular}
}
\label{tab:scalability}
\end{table}
\section{Countermeasures}
With the increasing focus on jailbreaking and red-teaming LLMs, recent efforts have turned towards defending against such attacks. Nevertheless, traditional prompt-based defenses \cite{robey2023smoothllm, cao2023defending, deng2023attack} are ineffective against activation attacks as they operate in the activation space without altering the prompt.
To defend against activation attacks, two approaches can be explored. 
Since steering vectors must be injected, the first approach involves utilizing a model checker to ensure that LLMs deployed for real-world use are clean and do not contain any additional files. 
The second approach involves investigating the implementation of a defense mechanism within the model itself. Ensuring that the addition of intermediate results disrupts the residual stream and does not generate meaningful output could provide a robust defense against activation attacks.
\section{Related Work}

\subsection{Alignment of LLMs}
Large Language Models (LLMs) have demonstrated exceptional performance across numerous tasks \cite{min2021recent, wei2022emergent, wei2022chain} and have been widely adopted for various applications. These LLMs can be categorized into foundational models like GPT-3 and instruction-tuned models like GPT-4. Foundational models are pre-trained on a vast textual corpus with objectives to predict subsequent tokens. This process can lead to behaviors that may diverge from social norms, ethical standards, and regulations. In contrast, instruction-tuned LLMs undergo additional fine-tuning to better align with user-provided instructions. A growing body of work on alignment \cite{wang2023aligning, liu2023trustworthy} including Supervised Fine-Tuning (SFT) of LLMs using annotated human instructions \cite{wang2022self} and Reinforcement Learning from Human Feedback (RLHF) \cite{ouyang2022training}. In this work, we investigate the vulnerability of aligned large language models to see if we can bypass their intended safeguards and provoke undesirable behaviors.

\subsection{Activation Engineering}
The word activation engineering was coined by \cite{turner2023activation}, as opposed to prompt engineering. Similar to prompts being used to control the output of LLMs, activation engineering involves adjusting the activation of individual layers during inference to modify the behavior of the models. This emerging area is pioneered by works studying the mechanistic interpretability of deep neural networks \cite{belinkov-etal-2017-neural, adi2016fine, belinkov-2022-probing, vig2020investigating, finlayson-etal-2021-causal}. To locate and modify the factual associations within autoregressive transformers, \cite{meng2022locating} developed a causal intervention method to identify neuron activations that are decisive in a model's factual predictions. They then proposed a method called ROME to modify feed-forward weights to update specific factual associations. Additionally, \cite{meng2022locating} finds that the MLP layer corresponds to the late site and attention corresponds to the early site. MLP layers are the ones that recall knowledge. \cite{meng2022mass} further improved upon the scalability of ROME to edit LMs' memories at a large scale. In contrast to weight editing methods, \cite{li2022emergent, li2023inference, hernandez2023measuring} proposed activation editing methods that apply steering vectors \cite{subramani2022extracting} to modify activations at inference time in order to alter model behavior. Our work builds upon the activation engineering \cite{turner2023activation} technique, where we reduce the need for manually selecting the intervention layer by performing contrastive layer search.

\subsection{Attack Large Language Models}
As LLMs become widely used in a variety of tasks, their robustness against adversarial attacks has drawn increased attention. Given the large size of LLMs, previous works on attacking language models \cite{zhang2020adversarial, li2020bert, deng2021tag, huang2023training, dong2023investigating} do not transfer well to attacks on LLMs. In the line of attacking LLMs, \cite{wan2023poisoning} shows that adversaries can poison instruction tuning datasets to systematically influence LLMs behavior. \cite{perez2022ignore} demonstrated that GPT-3 can be easily misaligned by simple handcrafted inputs. \cite{carlini2021extracting} proposed a training data extraction attack to recover individual training examples by querying the language model. \cite{du2022ppt} proposed a poisoned prompt tuning method called PPT. \cite{zou2023universal} proposed a method to automatically generate adversarial prompts. \cite{greshake2023not} proposed indirect prompt injection to strategically inject prompts into data likely to be retrieved.
\section{Conclusion}
In this paper, we comprehensively examine activation attacks on four safety alignments of LLMs. In particular, we propose a trojan activation attack framework called TA$^2$ that can effectively and efficiently break down the safeguards of LLMs, by injecting trojan steering vectors that can be triggered at inference time without being easily detected. Our experimentation across four prevalent alignments reveals that our attack framework is applicable to instruction-tuned LLMs of various sizes, demonstrating its efficacy while maintaining a lightweight profile. Additionally, we discuss potential defense mechanisms against activation attacks. We hope our work can draw attention to the potential vulnerabilities of aligned LLMs when subjected to activation attacks, thereby encouraging further research to enhance the robustness of LLMs.

\begin{acks}
This material is based upon work supported by the U.S. Department of Homeland Security under Grant Award Number 17STQAC00001-07-04, NSF awards (SaTC-2241068, IIS-2339198, and POSE-2346158), a Cisco Research Award, and a Microsoft Accelerate Foundation Models Research Award. The views and conclusions contained in this document are those of the authors and should not be interpreted as necessarily representing the official policies, either expressed or implied, of the U.S. Department of Homeland Security.
\end{acks}

\bibliographystyle{ACM-Reference-Format}
\balance
\bibliography{main}

\end{document}